# Coherent transport in a linear triple quantum dot made from a pure-phase InAs nanowire

Ji-Yin Wang,[†] Shaoyun Huang,[*,†] Guang-Yao Huang,[†] Dong Pan,[‡] Jianhua Zhao,[‡] and H. Q. Xu[*,†,§]

[†] Beijing Key Laboratory of Quantum Devices, Key Laboratory for the Physics and Chemistry of Nanodevices, and Department of Electronics, Peking University, Beijing 100871, China

[‡] State Key Laboratory of Superlattices and Microstructures, Institute of Semiconductors, Chinese Academy of Sciences, Beijing 100083, China

[§] Division of Solid State Physics, Lund University, Box 118, S-22100 Lund, Sweden

**Abstract**

A highly tunable linear triple quantum dot (TQD) device is realized in a single-crystalline pure-phase InAs nanowire using a local finger gate technique. The electrical measurements show that the charge stability diagram of the TQD can be represented by three kinds of current lines of different slopes and a simulation performed based on a capacitance matrix model confirms the experiment. We show that each current line observable in the charge stability diagram is associated with a case where a QD is on resonance with the Fermi level of the source and drain reservoirs. At a triple point where two current lines of different slopes move together but show anti-crossing, two QDs are on resonance with the Fermi level of the reservoirs. We demonstrate that an energetically degenerated quadruple point, at which all three QDs are on resonance with the Fermi level of the reservoirs, can be built by moving two separated triple points together via sophisticly tuning of energy levels in the three QDs. We also demonstrate the achievement of direct coherent electron transfer between the two remote QDs in the TQD, realizing a long-distance coherent quantum bus operation. Such a long-distance coherent coupling could be used to investigate coherent spin teleportation and super-exchange effects and to construct a spin qubit with an improved long coherent time and with spin state detection solely by sensing the charge states.

Corresponding authors: Professor Hongqi Xu (hqxu@pku.edu.cn) and Dr. Shaoyun Huang (syhuang@pku.edu.cn)
Beijing Key Laboratory of Quantum Devices, Key Laboratory for the Physics and Chemistry of Nanodevices, and Department of Electronics, Peking University, Beijing 100871, China

Quantum dots (QDs) made from narrow bandgap semiconductor nanowires are among the most promising building blocks for the physical implementation of electron spin based quantum information processing.[1-3] The advantages of using these narrow bandgap semiconductor materials arise from their small effective masses, strong spin-orbit interaction and large Landé g factors.[4-17] Coherent manipulations of single-spin states[18, 19] and two-spin states[20] in single QDs and double quantum dots (DQDs) have been demonstrated. Scaling up to triple quantum dots (TQDs) not only is a natural step of developments towards building up an integrated quantum system but also has advantages in resistance of decoherence,[21] all-electrical operation[22] and efficient spin teleportation.[23] The TQDs arranged in a linear[24-29] and a triangular configuration[30, 31] have been realized in two-dimensional electron gas (2DEG) systems formed in GaAs/AlGaAs heterostructures using top-gate techniques.

In this letter, we report on a highly tunable TQD, in a naturally linear configuration, realized in a single InAs nanowire by a local finger gate technique. The characteristics of the device are studied by electron transport measurements. The charge stability diagram of the TQD is extracted and is analyzed based on a capacitance matrix model. The results show that the controllability on inter-dot tunnel coupling, essential for exchange-based quantum gates,[22] has been achieved in the TQD device. We demonstrate successfully tuning of the device to energetically level-degenerated points of the three dots, i.e., the quadruple points (QPs), at which coherent states of the TQD can be formed. We further demonstrate the observation of coherent electron transport between the two remote dots with the middle dot in the well-defined Coulomb blockade regime.

The device is made from a single-crystalline pure-phase InAs nanowire with a diameter of ~35 nm on a Si/SiO$_2$ substrate. The nanowire is grown on a Si(111) substrate by Ag assisted molecular-beam epitaxy (MBE) and is of ~3-5 µm in length.[32, 33] Figure 1a shows a scanning electron microscope (SEM) image of the fabricated InAs nanowire device. The device consists of seven metal finger gates, labeled as G1 to G7 from left to right, with an average width of 30 nm and a pinch of 80 nm. Details about the InAs nanowire growth and the device fabrication are described in Supporting Information. Transport measurements are performed in a dilution refrigerator at a temperature of 40 mK. A bias voltage is applied to the source with the drain being grounded (see Figure 1b). Each local finger gate is capable of suppressing the channel current completely at a threshold voltage of -3 to -2.5 V (see details in Table S1 in Supporting Information), indicating that all the finger gates have a similar capacitive coupling to the nanowire. Here, we note that the back gate and uninvolved finger gates are grounded during the characterization measurements. When more than one finger gates are employed, the pinch-off threshold voltage of a finger gate may slightly deviate from the value determined above owing to gate cross-talk and screening effects. The Coulomb blockade and quantum confinement effects in a single QD defined in the InAs nanowire by an axial double barrier using, e.g., local finger gates G5 and G7, can be readily evaluated by measuring the charge stability diagram (see Part III in Supporting Information). The electron additional energy and quantization energy extracted from the measured charge stability diagram is ~13.5 and ~2 meV, respectively. Different multiple-QDs can be created in the InAs nanowire by properly designed combinations of voltages applied to the local finger gates.

Figure 2 shows the charge stability diagrams of two DQDs defined by local finger gates in the nanowire as shown in the cross-sectional schematics of the figure. Figure 2a shows the measured charge stability diagrams of the left DQD (DQD1) defined with finger gates G1, G3 and G5 as schematically shown in Figure 2b. Here, the tunneling barrier between source (drain) and the left (right) dot is generated by voltage $V_{G1}$ ($V_{G5}$) applied to gate G1 (G5). The inter-dot coupling strength is controlled by gate G3, and the source-drain current $I_{SD}$ is measured as a function of voltages $V_{G2}$ and $V_{G4}$ applied to finger gates G2 and G4 at a source-drain bias voltage of $V_{SD}$=35 µV. Figure 2c shows the corresponding charge stability diagrams measured for the right DQD (DQD2) defined with finger gates G3, G5 and G7 as schematically shown in Figure 2d, i.e., the source-drain current $I_{SD}$ as a function of voltages $V_{G4}$ and $V_{G6}$ applied to finger gates G4 and G6 at $V_{SD}$=35 µV. In Figures 2a and 2c, finite current flows are observed not only at the corners (triple points) of hexagons but also at the boundaries of the hexagons. The current flow at the boundaries of the hexagons is caused by co-tunneling processes when any one of the two dots is resonant in energy with the Fermi level of the source and drain reservoirs.[34] Sharply defined honeycomb-shaped structures reveal that the two DQDs are well defined and are set in the weak inter-dot tunnel coupling regime. The linear transport characteristics of a DQD in the weak coupling regime can be described by a mutually coupled capacitance model.[34] From the boundaries of the hexagon regions marked by the dashed squares in Figures 2a and 2c, we can extract the capacitances and the single-electron charging energies for the two DQDs (see Table S2 in Supporting Information). For example, the single-electron charging energies of the two individual QDs in the DQD1 are extracted to be 7.5 and 8.4 meV while the charging energies of the two individual QDs in DQD2 are 10.7 and 11.6 meV, which are all on the similar order of magnitude as found for the charging energy of a single QD in the device. Figures 2a and 2c clearly show that the two DQDs exhibit approximately the same transport characteristics, demonstrating an excellent controllability and the uniformity of the local finger gates in the device. The high functionality of the local finger gates allows constructing and manipulating a highly tunable TQD in the pure-phase InAs nanowire.

Figure 3 shows the experimental and simulated transport characteristics of a TQD defined in the InAs nanowire using gates G1 to G7. Gate G1 (G7) is used to generate an outer tunneling barrier and to manipulate the tunneling transparency between the source (drain) and the leftmost (rightmost) dot. Gates G3 and G5 are used to control the inter-dot tunnel coupling between the leftmost dot (QD1) and the middle dot (QD2), and between the rightmost dot (QD3) and the middle dot (QD2), respectively. Gates G2, G4 and G6 are used to tune the electrostatic potential of QD1, QD2 and QD3. Here, we note that gates G3 and G5 not only generate tunneling barriers right above the gates but also laterally couple to the two neighboring dots via parasitical capacitances. Thus, in order to map out the transport characteristics of the TQD more efficiently, we have selected one barrier gate (G3) and one plunger gate (G6) to tune the TQD in the measurements shown in Figure 3 (and also in the following measurements shown in Figure 4). In the sequential tunneling regime with opaque barriers, an electron can transport from the source to the drain through the TQD mainly at a quadruple point, where all the three dots are on resonance and four involved charge occupation states are energetically degenerated.[35] However, a finite transport current can be also observed in a wide range of $V_{G3}$ and $V_{G6}$ in Figure 3. The existence of the finite transport current derives from co-tunneling processes when any one of the three dots is resonant in energy with the Fermi level of the source and drain reservoirs.[36] Figure 3a shows the source-drain current $I_{SD}$ of the TQD measured as a function of gate voltages $V_{G3}$ and $V_{G6}$ at $V_{SD}$=35 μV in a relatively strong coupling region. The measured tunneling current peaks form the current lines with three different slopes, nearly horizontal, tilted and nearly vertical current lines, as illustrated by a dashed, a dot-dashed, and a dotted line, respectively, in Figure 3a. Taking into the dot position and the gate location account, we can attribute the nearly horizontal, tilted, and nearly vertical current lines to the co-tunneling when QD1, QD2, and QD3 are on resonance with the Fermi level of the reservoirs, respectively. Whenever any two of the resonant current lines move closer, an avoided crossing happens due to tunnel coupling between the levels in the two corresponding dots.[37] The inter-separated distance in the avoided crossing region reflects the inter-dot coupling strength. The capability to manipulate the inter-dot tunnel coupling is essential for establishing exchange-based spin qubits in the TQD.[22, 38] In Figure 3c, we demonstrate such capability by tuning the gate voltage $V_{G5}$ from -2.65 V in Figure 3a to -2.75 V. In contrast to Figure 3a, the bending angle in the avoided crossing region marked by green arrows becomes sharper, indicating that the tunnel coupling between QD2 and QD3 becomes weaker. Furthermore, the inter-separated distance marked by the green arrows is decreased [cf. Figures 3b and 3d], implying a drop in the capacitive coupling between QD2 and QD3. The similar behaviors are found in the avoided crossing region marked by the gray arrows. Thus, tuning the gate voltage $V_{G5}$ to a more negative value can also lead to drops in the capacitive coupling and in the tunnel coupling between QD1 and QD3 as expected. The prominent features of experimental charge stability diagrams shown in Figures 3a and 3c can be well reproduced by simulations using a capacitance matrix model,[35, 36] see the results shown in Figures 3b and 3d (for the details of the simulations, see Part V in Supporting Information). Indeed a highly tunable TQD can be constructed with the local finger gates in our single InAs nanowire device and can help us to access some specific coupling regimes, such as energetically degenerate quadruple points (QPs), which has not been reported before.

Figure 4 shows how a QP can be realized from two separated triple points by tuning the electrostatic potential of QD2 with gate G4 in the TQD device. Here, the TQD formed is the same as in Figure 3, but has been set to the case with a stronger inter-dot tunnel coupling. In the strong inter-dot tunnel coupling regime, triple points can still be located by extending straight boundary current lines in the charge stability diagram.[34] Figure 4a shows the charge stability diagram of the TQD at $V_{G4}=$ -0.048 V. The dashed, dot-dashed, and dotted lines represent current lines arising from co-tunneling processes in the cases when QD1, QD2, and QD3 are on resonance with the Fermi level of the source and drain reservoirs, respectively. At the triple point marked by a yellow circle, QD2 and QD3 are on resonance with the Fermi level of the source and drain reservoirs. At the triple point marked by a white circle, QD1 and QD3 are on resonance with the Fermi level of the reservoirs. However, at the triple point marked by the white circle, QD2 is generally detuned from QD1 and QD3. Thus, in this case, gate G4, the plunger gate for QD2, is selected to tune QD2 back to resonance with QD1 and QD3. At $V_{G4}$=-0.045 V (Figure 4b), the positions of the two triple points marked by the yellow and white circles move closer as compared with those in Figure 4a. Increasing $V_{G4}$ mainly causes energy levels of QD2 to drop, but it also leads to moving down the energy levels of QD3 owing to the presence of a cross-talk capacitance between gate G4 and QD3. Furthermore, the energy levels of QD1 almost do not move with increasing $V_{G4}$, revealing the presence of a weak cross-talk capacitive coupling between gate G4 and QD1. As a consequence, the position of the triple point marked by the yellow circle moves toward to more negative values of both $V_{G3}$ and $V_{G6}$, while the position of the triple point marked by the white circle moves toward to more negative values of $V_{G6}$ mainly. As a result, the positions of the two triple points get closer in the charge stability diagram with increasing $V_{G4}$. Figure 4c shows the charge stability diagram of the TQD at $V_{G4}$=-0.04 V. Here, the electron levels of QD2 and QD3 move down further as $V_{G4}$ is increased and the two marked triple points get even closer. Finally, as shown in Figure 4d, with further increasing $V_{G4}$ to $V_{G4}$=-0.03 V, the two triple points are merged and a QP marked by the green circle is formed in the TQD. Now, three electron levels (with one from each dot) of the three QDs are aligned in energy, forming a coherent, resonant state. Note that the alignment of the electron levels from different QDs is an ingredient step for the quantum cellular automata (QCA) process in a multiple-QD device.[39] The triple points marked by the yellow arrows in Figure 4c are in similar situations as the one marked by the white circle in the figure, where QD1 and QD3 are on resonance with the Fermi level of the source and drain reservoirs. At all these triple points, direct electron transfers between the two remote dots (QD1 and QD3) can take place, as we shall show below.

Figure 5a shows the measured $I_{SD}$ of the TQD as a function of $V_{G2}$ and $V_{G6}$ at $V_{SD}$=35 μV in a less inter-dot coupling strength region, realized by properly setting the voltages applied to the other five finger gates in the device. The nearly horizontal (vertical) lines correspond to the situations in which a QD1 (QD3) level is on resonance with the Fermi level of the reservoirs. Here, no resonant current lines derived from the QD2 levels are observable, indicating that the QD2 is in the Coulomb blockade regime. Thus, at an energetically degenerated triple point, direct electron transfer between the two remote dots (QD1 and QD3) occurs. Such energetically degenerated triple points at which direct electron transfer processes can take place are seen in a wide voltage region of $V_{G2}$ and $V_{G6}$ in Figure 5a. To see more clearly the direct electron transfers between the two remote QDs in the TQD, we show the charge occupation configurations (M, N, P) in the Coulomb blockade regions surrounding the blue point A in Figure 5a, where M, N and P are the electron occupation numbers in QD1, QD2 and QD3, respectively. Figure 5b is the zoom-in plot in the region marked by the green rectangle in Figure 5a together with the measured current $I_{SD}$ along a line cut indicated in the figure. The blue point A is located at the boundary between the (M+1, N, P) and (M, N, P+1) regions. When crossing through the blue point A from the (M+1, N, P) region to the (M, N, P+1) region, an electron is transferred directly from QD1 to QD3. At the blue point A, the TQD is in a coherent state with an equal probability for the last occupied electron to be found in QD1 or QD3. A clear current peak against the noise floor at point A can be seen and can be attributed to resonant tunneling through the coherent state.[29] This state could be used for spin bussing and for construction of a spin qubit. Since, fundamentally, the state can be considered as a three-electron state, with two electrons forming a singlet in one of two outer QDs and the remaining electron forming a doublet in the other outer QD. Thus, the charge transfer between the two outer QDs is necessarily accompanied by a transfer of spin. An alternative interpretation is to attribute the current to co-tunneling via virtual occupations of QD2, i.e., (M, N+1, P) and (M+1, N-1, P+1) charge state,[28, 29, 38] as shown in Figures 5c and 5d. Here, electrons tunnel through the TQD via the initial (M+1, N, P) charge state and final (M, N, P+1) charge state. In Figure 5c, an electron indicated by a solid dot tunnels out of QD1, through virtual occupation of the $\mu_2$(M, N+1, P) level of QD2, to QD3. In Figure 5d, the tunneling process could be better described by hole transport, i.e., a hole indicated by an open dot tunnels out of QD3, through virtual occupation of the $\mu_2$(M+1, N-1, P+1) level of QD2, to QD1. In terms of electron transport, the process described in Figure 5d is equivalent to the process that an electron in QD2 tunnels into QD3 and in the same time an electron in QD1 tunnels into QD2. The processes described in Figures 5c and 5d are coherent transfers of an electron from QD1 to QD3 and the tunneling current observed at blue point A is a coherent sum up of all such individual transfer processes. Finally, we would like to note that the observation of such a long-distance coherent electron transfer[29, 40] is an important step towards the realization of coherent spin-qubit teleportation[23] and the study of super-exchange interaction processes.[22]

In conclusion, a highly tunable linear TQD has been realized in a single-crystalline pure-phase InAs nanowire using a local finger gate technique. The TQD has been analyzed by electrical measurements. The measured charge stability diagram of the TQD is characterized by three groups of current lines with three different slopes. A simulation based on a capacitance matrix model has been carried out for the TQD and the results are fully in agreement with the measurements. It has been identified that each current line is associated with a case that a QD is on resonance with Fermi level of the source and drain reservoirs. At a triple point where two current lines of different slopes move together but show anti-crossing, two QDs are on resonance with the Fermi level of the reservoirs. We have demonstrated that the inter-dot tunnel coupling is tunable in our TQD device. An efficient manipulation of inter-dot coupling is essential for the operation of exchange-based spin qubits. We have also shown that the energetically degenerated quadruple point, at which all three QDs are on resonance with the Fermi level of the reservoirs, can be achieved from moving two separated triple points together via sophistically tuning of the energy levels in the three QDs by using the local finger gates. The formation of a quadruple point in a multiple QD system is on demand for the QCA operation. We have solidly demonstrated that it is possible to achieve direct coherent electron transfer between the two remote QDs in the TQD. Such a long-distance coherent quantum bus operation is desired for the investigations of coherent spin teleportation and super-exchange interaction in a multiple QD system. Finally, we note that our highly tunable linear TQD system could be used as a versatile platform to investigate virtual state-assistant spin-dependent coherent tunneling processes between remote QDs[41, 42] and to construct spin qubits with a long coherent time and with spin state detection solely by sensing the charge states.

**ACKNOWLEDGEMENTS**

This work is supported by the Ministry of Science and Technology of China through the National Key Research and Development Program of China (Grant Nos. 2016YFA0300601 and 2016YFA0300800), and the National Natural Science Foundation of China (Grant Nos. 91221202, 91421303, 11274021 and 61504133). HQX also acknowledges financial support from the Swedish Research Council (VR).

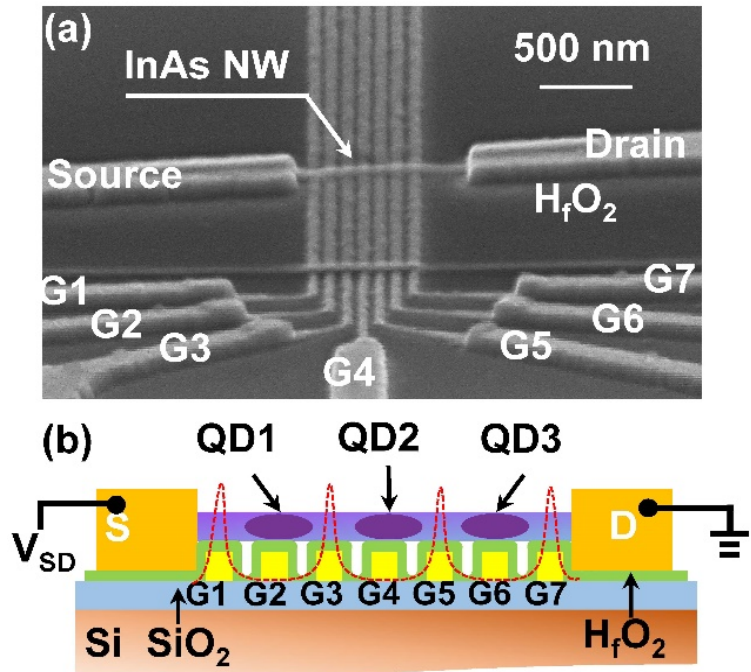

**Figure** 1. (a) Scanning electron microscope image of an InAs nanowire device investigated in this work. The nanowire has a diameter of ~35 nm. The array of seven finger gates are fabricated below the InAs nanowire. The finger gates are made from Ti/Au (5/15 nm in thickness) and have a width of ~30 nm. The array has a pitch of 80 nm and are isolated from the nanowire by a 10-nm-thick $HfO_2$ film. (b) Cross-sectional illustration of the device structure. The TQD is defined and manipulated using finger gates G1 to G7.

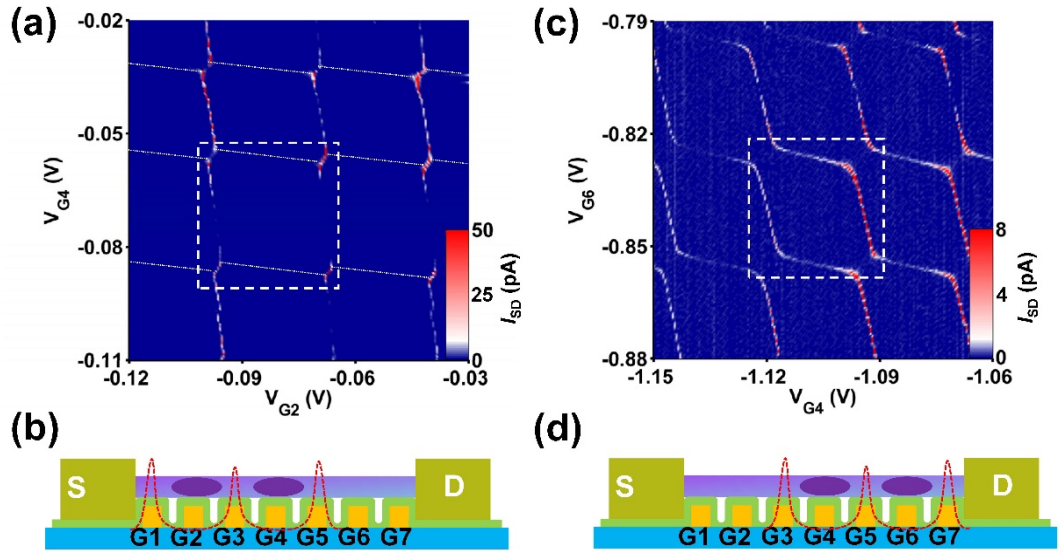

**Figure** 2. (a) Charge stability diagram of the left DQD defined by using gates G1, G3 and G5. The measurements are performed at $V_{SD}$= 35 µV, with the gate voltage $V_{G1}$, $V_{G3}$ and $V_{G5}$ being set at -3.3, -2.35 and -2.8 V, respectively. (b) Schematic illustration of the left DQD. (c) Charge stability diagram of the right DQD defined by using gates G3, G5 and G7. The measurements are performed at $V_{SD}$= 35 µV, with the gate voltage $V_{G3}$, $V_{G5}$ and $V_{G7}$ being set at -2.55, -2.6 and -2.3 V, respectively. (d) Schematic illustration of the right DQD.

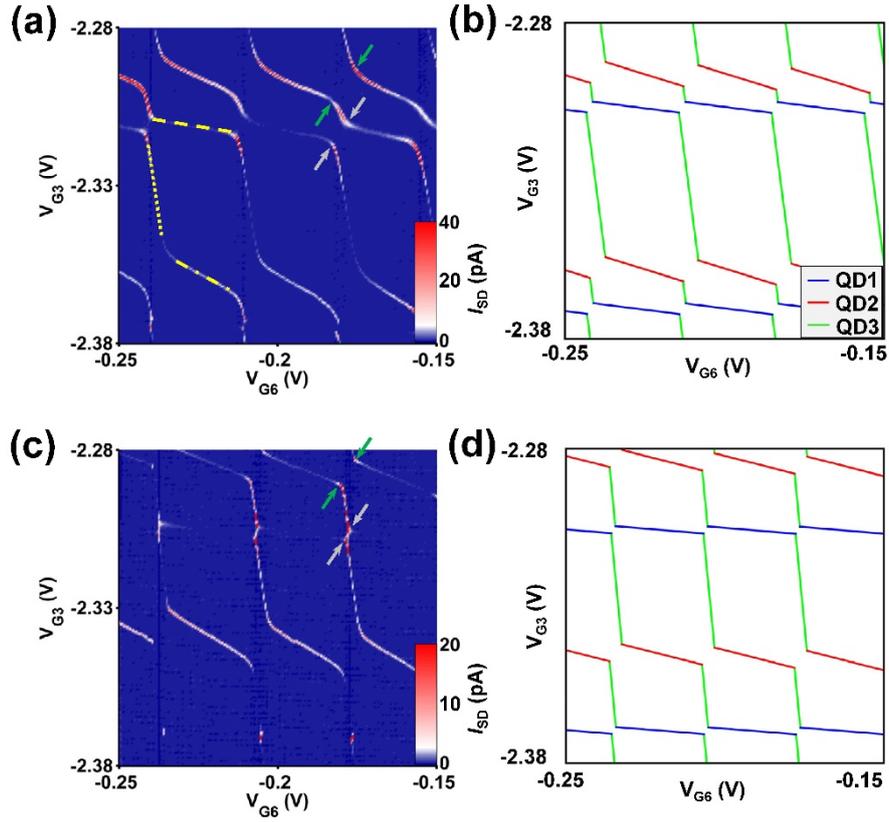

**Figure** 3. Source-drain current $I_{SD}$ of the TQD measured as a function of gate voltages $V_{G3}$ and $V_{G6}$ at $V_{SD}$ = 35 μV. The TQD is defined by setting the two outer barrier gates at $V_{G1}$ = -3.3 V and $V_{G7}$ = -2.45 V, the plunger gates at $V_{G2}$ = 0 V and $V_{G4}$ = -0.08 V, and the inter-dot coupling gate at $V_{G5}$= -2.65 V (a) and at $V_{G5}$= -2.75 V (c). (b) and (d) Simulated stability diagrams based on a capacitance matrix model (see details in Part V in Supporting Information). (b) shows the simulation results for the measurements shown in (a) and (d) shows the simulation results for the measurements shown in (c).

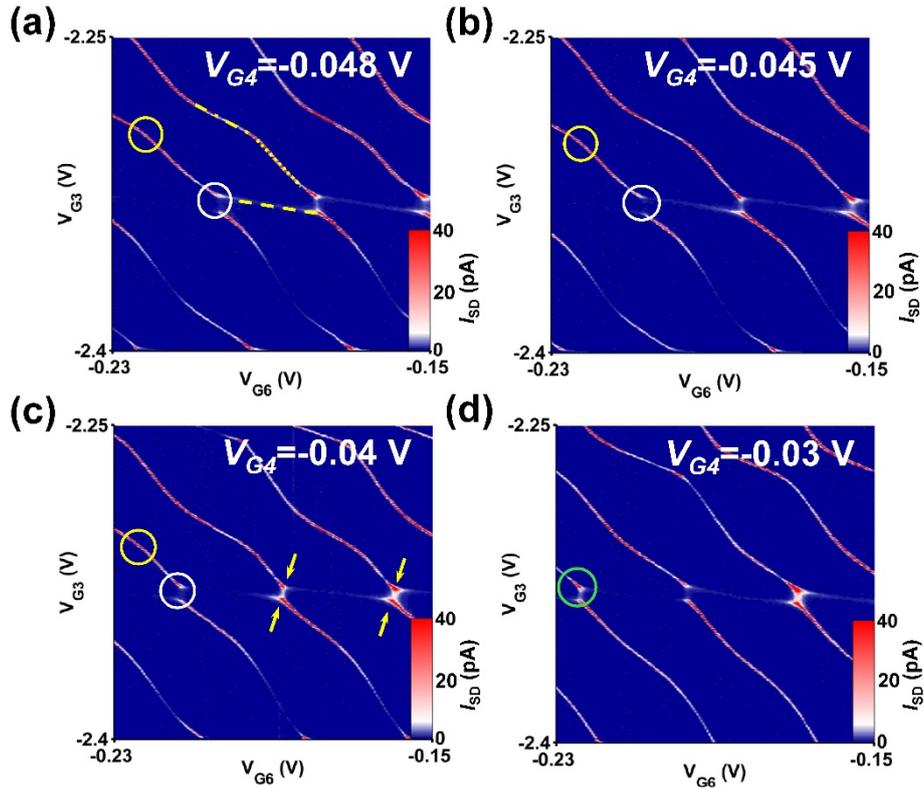

**Figure** 4. Stability diagrams of the TQD measured at $V_{SD}$ = 35 μV and at $V_{G4}$= -0.048 V (a), $V_{G4}$= -0.045 V (b), $V_{G4}$= -0.04 V (c), and $V_{G4}$= -0.03 V (d). In the measurement, gate voltages $V_{G1}$, $V_{G2}$, $V_{G5}$ and $V_{G7}$ are set at $V_{G1}$ = -3.2 V, $V_{G2}$ = 0 V, $V_{G5}$ = -2.55 V, and $V_{G7}$ = -2.45 V. The triple point marked by the yellow circle corresponds to the case that both QD2 and QD3 are on resonance with the Fermi level of the reservoirs. The triple point marked by the white circle corresponds to the case that both QD1 and QD3 are on resonance with the Fermi level of the reservoirs. The green circle marks the QP of the TQD at which QD1, QD2, and QD3 are all on resonance with the Fermi level of the reservoirs.

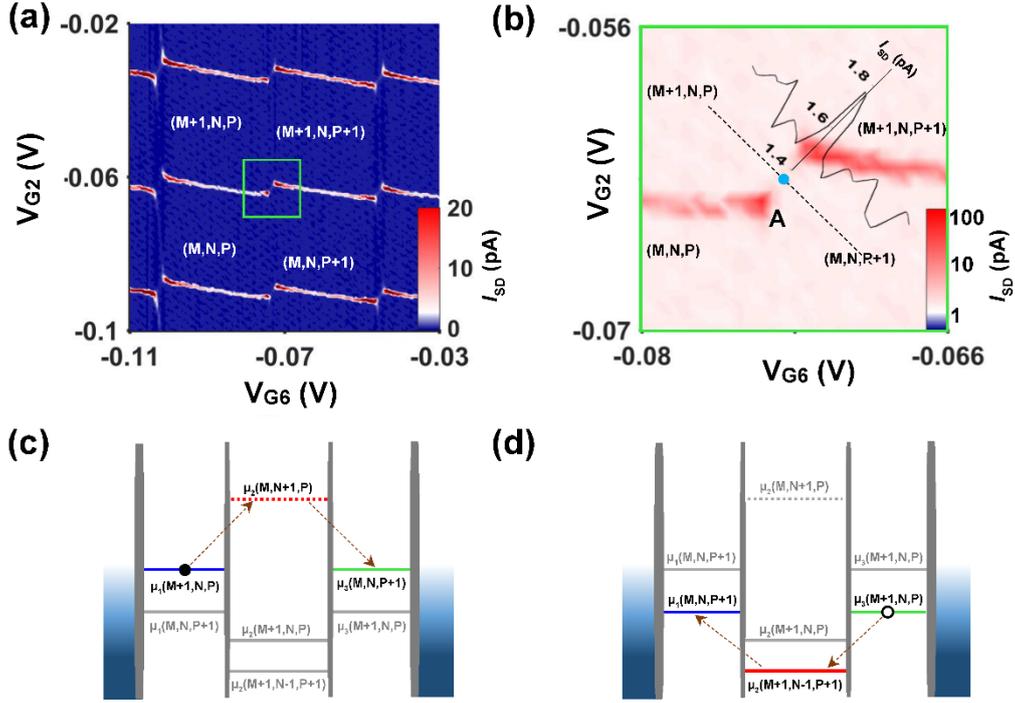

**Figure** 5. (a) $I_{SD}$ as a function of $V_{G2}$ and $V_{G6}$ measured for the TQD at $V_{SD}$ = 35 μV. The other gate voltages are set at $V_{G1}$ =-3.3 V, $V_{G3}$ = -2.4 V, $V_{G4}$ = 0 V, $V_{G5}$ = -2.6 V, and $V_{G7}$ = -2.45 V. Here, the charge configurations in different Coulomb blockade regions are indicated as (M, N, P), where M, N, and P are occupation numbers of electrons in QD1, QD2, and QD3, respectively. (b) Zoom-in plot of the measurements in the region marked by the green rectangle in (a) together with the measured current $I_{SD}$ along the dashed line shown in the panel. The blue point A corresponds to the case that a level of QD1 and a level of QD3 are aligned and on resonance with the Fermi level of the reservoirs. (c) and (d) Energy diagrams in the TQD and possible co-tunneling processes between the two outer QDs from the (M+1, N, P) to (M, N, P+1) charge state via virtual occupations of a (M, N+1, P) charge state and a (M+1, N-1, P+1) charge state of the TQD. Here, the solid dot represents the co-tunneling process carried by electrons and the open dot the co-tunneling process by holes.
``

**Supporting Information for**

**Coherent transport in a linear triple quantum dot made from a pure-phase InAs nanowire**

Ji-Yin Wang,[†] Shaoyun Huang,[*,†] Guang-Yao Huang,[†] Dong Pan,[‡] Jianhua Zhao,[‡] and H. Q. Xu[*,†,§]

[†] Beijing Key Laboratory of Quantum Devices, Key Laboratory for the Physics and Chemistry of Nanodevices, and Department of Electronics, Peking University, Beijing 100871, China

[‡] State Key Laboratory of Superlattices and Microstructures, Institute of Semiconductors, Chinese Academy of Sciences, Beijing 100083, China

[§] Division of Solid State Physics, Lund University, Box 118, S-22100 Lund, Sweden

*Emails: hqxu@pku.edu.cn; syhuang@pku.edu.cn

**Abstract**

In this Supporting Information, we provide the details of the InAs nanowire employed in the main article, device fabrication process and physical parameters extracted from the measurements. A single quantum dot (QD) defined in a single-crystalline pure-phase InAs nanowire is investigated and this building block is used for the construction of multiple QDs. Furthermore, a capacitance matrix model is adopted to simulate the electron transport characteristics of triple quantum dot (TQD) in the linear response regime.

## I. InAs nanowire and device fabrication

The InAs nanowire employed in the device fabrication for this work are among those grown on Si (111) substrates by molecular-beam epitaxy (MBE) using Ag catalysts.[1,2] The grown InAs nanowires have a length of ~3-5 µm and a diameter of 10-50 nm. It has been found that the InAs nanowires grown by MBE with small diameters are in pure phase and are either single wurtzite or single zincblende crystals, and free from stack faults and extended defects.[2] Figure S1(a) shows the transmission electron microscope (TEM) images of a representative InAs nanowire. The nanowire is grown along the $[12\bar{1}]$ direction and is a single zincblende crystal. Figure S1(b)-S1(d) shows the high-resolution transmission electron microscope (HRTEM) images of top, middle and bottom parts of the nanowire. We observe that the whole nanowire is a single crystal and is in pure phase, which ensures that we can define quantum dots (QDs) at any part of the nanowire without undesired barriers derived from stack faults and defects.

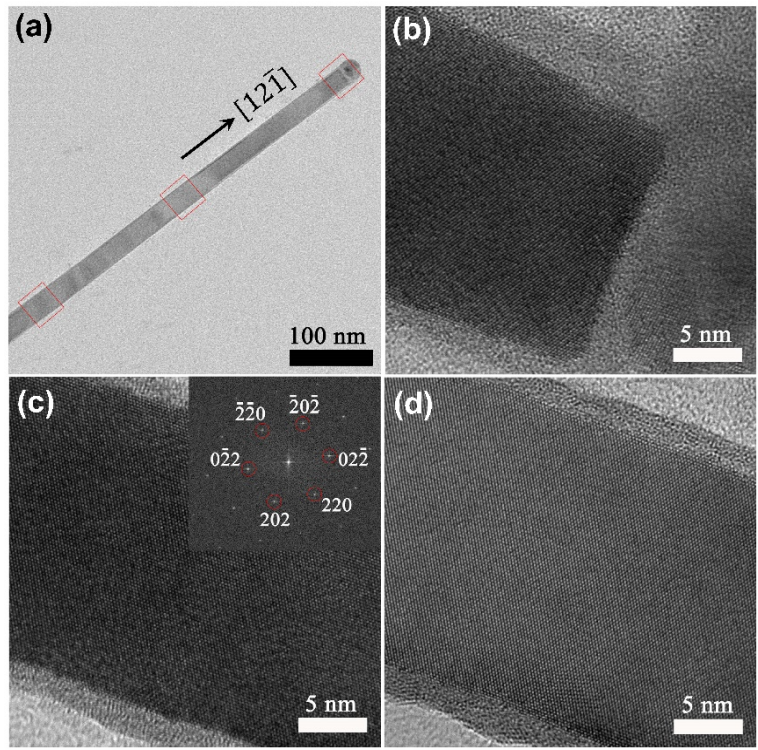

**Figure S1.** (a) TEM image of a representative single-crystalline pure zincblende-phase InAs nanowire with a diameter of 27 nm, grown along the $[12\bar{1}]$ direction. (b)-(d) HRTEM images of the top, middle and bottom parts of the InAs nanowire. Inset in (c) shows the fast Fourier transform (FFT) spectrum obtained from the HRTEM image of the middle segment of the nanowire.

With such a single-crystalline pure-phase InAs nanowire, we are able to fabricate a local finger gate device as shown in Figure 1a of the main article. The device fabrication begins from preparation of finger gate arrays on a silicon substrate covered with a 200-nm-thick layer of $SiO_2$. Each array contains seven finger gates, labeled as G1 to G7 from left to right (see Figure 1 in the main article). Electron-beam lithography (EBL) is used to define patterns of the finger gates on PMMA resist and electron-beam evaporation (EBE) is used to deposit 5-nm-thick titanium and 15-nm-thick gold. After lift-off process, the finger gates with an averaged width of 30 nm and a pitch of 80 nm are obtained. Subsequently, the finger gate arrays are covered by a layer of $HfO_2$ with a thickness of 10 nm by means of atomic layer deposition (ALD). Then, InAs nanowires are transferred from the growth substrate onto the finger gate arrays. The finger gate arrays each occupied by only one single nanowire with a diameter of ~30 nm are selected for final device fabrication. Here, the nanowires with a diameter of ~30 nm are selected, because such nanowires are thin enough to be in a pure phase and are in the same time thick enough to avoid a significant increase in the contact resistance arising from quantum confinement.[3] The source and drain contact regions on two ends of each nanowire are defined and opened again by EBL. The contact regions are then etched in diluted $(NH_4)_2S_x$ solution, in order to remove nanowire surface oxide and to achieve surface passivation, right prior to metal deposition.[4, 5] The device is completed with fabrication of contact electrodes by deposition of 5-nm-thick titanium and 90-nm-thick gold and lift-off process.

**II. Current-suppression threshold voltages of local figure gates**

The local finger gates are characterized individually by measurements of the source-drain current of the InAs nanowire as a function of a voltage applied to an individual finger gate at source-drain bias voltage $V_{SD}$=1 mV. In the measurements, all uninvolved finger gates and back gate are grounded. The threshold voltage of a finger gate, at which or below it the nanowire current can be completely suppressed.

**Table S1.** Threshold voltages of seven local finger gates in the device studied in the main article. The source-drain current can be completely suppressed at the threshold voltage, or a voltage below it, applied to a finger gate. In the characterization for a figure gate, all uninvolved finger gates and back gate are grounded, and a source-drain bias voltage of 1 mV is applied to the nanowire.

| Gate | G1 | G2 | G3 | G4 | G5 | G6 | G7 |
|---|---|---|---|---|---|---|---|
| **Threshold (V)** | −3.0 | −2.58 | −2.63 | −2.88 | −2.98 | −2.78 | −2.55 |

### III. Transport measurement of a single QD defined in the InAs nanowire

Figure S2 shows the differential conductance $dI_{SD}/dV_{SD}$ of a single quantum dot (QD) as a function of source-drain voltage $V_{SD}$ and gate voltage $V_{G6}$ (charge stability diagram). Here the QD is defined by gates G5 and G7. The electrostatic potential of the QD is tuned by voltage $V_{G6}$ applied to gate G6. Diamond-shaped Coulomb blockade regions are clearly observed in the charge stability diagram. The close points seen at zero bias voltage between neighboring Coulomb diamonds indicate that the transport is dominantly characterized by single electron tunneling through the gate-defined single QD. Conductance lines due to excited states in the QD are also clearly observed. From the measurements, the addition energy ($E_a$) and quantization energy ($\Delta E$) of the QD are extracted to be ~13.5 and ~2 meV, respectively. These single-QD features can be observed in a QD defined by applying similar voltages to any other three neighboring finger gates.

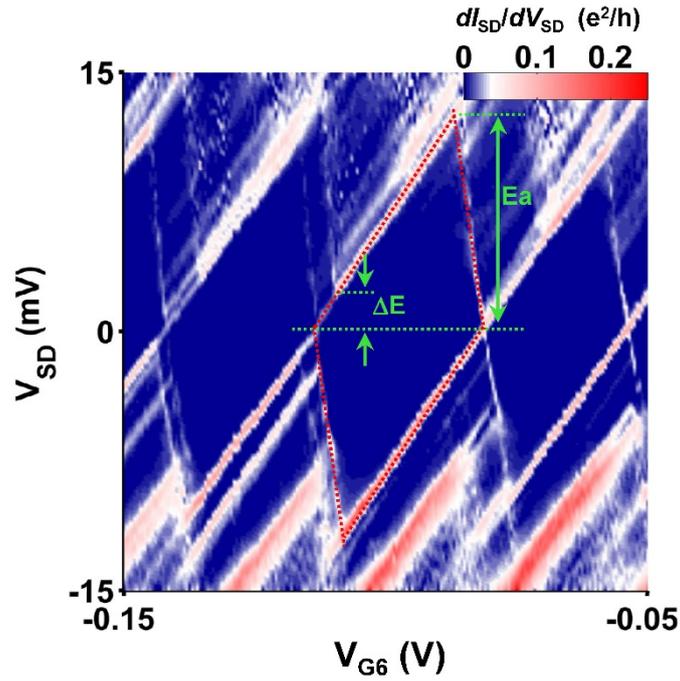

**Figure S2.** Source-drain current $I_{SD}$ measured for a single QD defined in the InAs nanowire by using gates G5 and G7 as a function of bias voltage $V_{SD}$ and gate voltage $V_{G6}$ (charge stability diagram). Here, the voltages applied to gates G5 and G7 to define the QD are $V_{G5} = -2.95$ V and $V_{G7} = -2.55$ V.

## IV. Capacitances extraction for DQDs defined in the InAs nanowire by local gates

The charge stabilities diagrams of two double quantum dots (DQDs) shown in Figure 2 of the main article are characterized by electrical measurements in the linear response regime. Well-defined hexagonal honeycomb-shape structures are observed in the charge stability diagrams of the two DQDs, which means that the DQDs are in the weak tunnel coupling regime. Linear response transport of a DQD in the weak tunnel coupling regime can be described with constant interaction (CI) model.[6] From the hexagon regions marked by the dashed squares in Figures 2a and 2c of the main article, the characteristic device parameters of the DQDs are extracted. The results are shown in Table S2.

**Table S2.** Extracted device parameters (gate capacitances to individual dots, coupling capacitance between dots, and charging energies of individual dots) of the two DQDs studied in the main article from the hexagons marked by the dashed squares in Figures 2a and 2c of the main article.

| DQD1 | $C_{G2\text{-}QD1}$ (aF) | $C_{QD1}$ (aF) | $E_{C,QD1}$ (meV) | $C_{G4\text{-}QD2}$ (aF) | $C_{QD2}$ (aF) | $E_{C,QD2}$ (meV) | $C_M$ (aF) | $C_{G2\text{-}QD2}$ (aF) | $C_{G4\text{-}QD1}$ (aF) |
|---|---|---|---|---|---|---|---|---|---|
| (G1-G5) | 5.4 | 21.3 | 7.5 | 5.3 | 19.1 | 8.4 | 1.6 | 0.2 | 0.3 |
| DQD2 | $C_{G4\text{-}QD2}$ (aF) | $C_{QD2}$ (aF) | $E_{C,QD2}$ (meV) | $C_{G6\text{-}QD3}$ (aF) | $C_{QD3}$ (aF) | $E_{C,QD3}$ (meV) | $C_M$ (aF) | $C_{G4\text{-}QD3}$ (aF) | $C_{G6\text{-}QD2}$ (aF) |
| (G3-G7) | 6.3 | 15 | 10.7 | 5.1 | 13.8 | 11.6 | 3.3 | <0.1 | 0.3 |

## V. Capacitance matrix model for the description of the TQD studied

The transport of the TQD studied in the main article in the linear response regime is modeled by the classical theory,[6] which is described as a network of tunnel resistors and capacitors. The equivalent circuit diagram is depicted in Figure S3, which includes the nodes of three QDs (QD1, QD2, and QD3), three gates (G3, G4, and G6), and source (S) and drain (D), as indicated by solid dots in the circuit diagram. In the capacitive matrix model, these nodes are assumed to be only capacitively coupled. The charge and electrostatic potential on each node are denoted as $Q_j$ and $V_j$ with subscript $j = 1$, 2, 3, G3, G4, G6, S, or D.

The charges and electrostatic potentials on the nodes are related by

$$\begin{bmatrix} Q_{QD} \\ Q_G \end{bmatrix} = \begin{bmatrix} C_{QD} & C_{QD,G} \\ C_{G,QD} & C_G \end{bmatrix} \begin{bmatrix} V_{QD} \\ V_G \end{bmatrix},$$

where $Q_{QD} = [Q_1 \ Q_2 \ Q_3]^T$, $V_{QD} = [V_1 \ V_2 \ V_3]^T$, and $V_G = [V_S \ V_{G3} \ V_{G4} \ V_{G6} \ 0]^T$

with $V_S = V_{SD}$ and $V_D = 0$, i.e., a fixed 0 of the electrostatic potential at the ground.

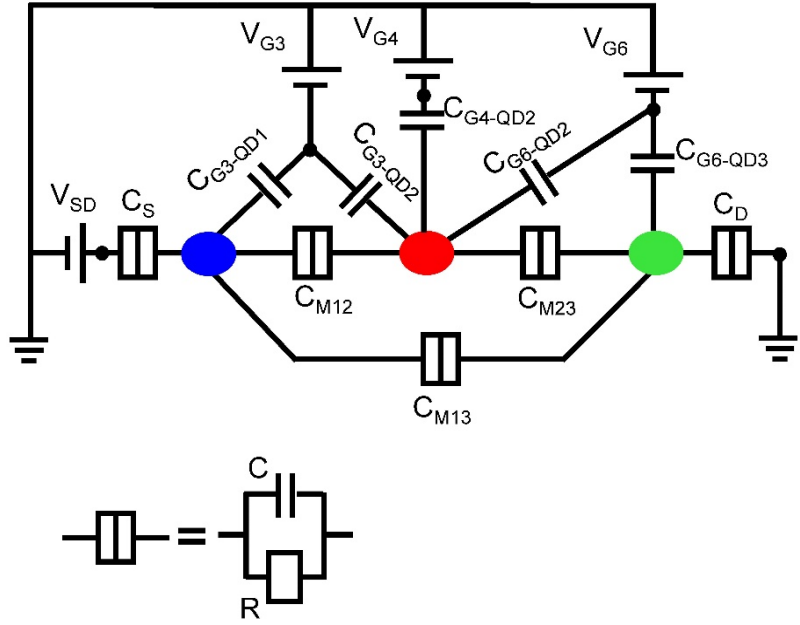

**Figure S3.** Electrostatic capacitance model including the most significant capacitances to simulate transport characteristics of the TQD studied in the main article. Gates G3 tunes the electron number in QD1 and QD2, gate G6 tunes the electron number in QD3 and QD2 and gate G4 only tunes the electron number in QD2.

The block capacitance matrix $C_{QD}$ connects the charges on the three QDs to the electrostatic potentials of the three QDs, while $C_{QD,G}$ connects the charges on the QDs to the electrostatic potentials of the gates and the source and drain. In the linear response regime, $V_{SD} \approx 0$. The total charge on each node $Q_i$ is the sum of the charge of all the capacitors connected to it,

$$Q_i = \sum_{k=1}^{8} C_{ik}(V_i - V_k) , \qquad (1)$$

where $C_{ik}$ is the capacitance between node $i$ and $k$, of which with nonzero value are shown in Figure S3. To find the electrostatic energy of the TQD, only the equations involving $Q_{QD}$ are needed,

$$Q_{QD} = C_{QD}V_{QD} + C_{QD,G}V_G ,$$

where the capacitance matrices $C_{QD}$ and $C_{QD,G}$ can be established using Eq. (1). Explicitly,

$$C_{QD} = \begin{bmatrix} C_1 & -C_{M12} & -C_{M13} \\ -C_{M12} & C_2 & -C_{M23} \\ -C_{M13} & -C_{M23} & C_3 \end{bmatrix} ,$$

and

$$\begin{aligned}C_1 &= C_{G3-QD1} + C_{M12} + C_{M13} + C_S , \\ C_2 &= C_{G3-QD2} + C_{G4-QD2} + C_{G6-QD2} + C_{M12} + C_{M23} , \\ C_3 &= C_{G6-QD3} + C_{M13} + C_{M23} + C_D ,\end{aligned}$$

and

$$C_{QD,G} = \begin{bmatrix} -C_S & -C_{G3-QD1} & 0 & 0 & 0 \\ 0 & -C_{G3-QD2} & -C_{G4-QD2} & -C_{G6-QD2} & 0 \\ 0 & 0 & 0 & -C_{G6-QD3} & -C_D \end{bmatrix} .$$

The electrostatic potentials on the QDs are

$$V_{QD} = C_{QD}^{-1}(Q_{QD} - C_{QD,G}V_G) ,$$

and, subsequently, the electrostatic energy of the TQD can be obtained from

$$E = \frac{1}{2}V_{QD}C_{QD}V_{QD} ,$$

where the charge on QD $i$ is $Q_i = -eN_i$, where $e = 1.6 \times 10^{-19}$ C. Evidently, the electrostatic energy of the TQD is a function of charge configuration $(N_1, N_2, N_3)$ in the TQD

The chemical potential for adding the $N_i$-th electron to QD $i$ is defined as

$$\mu_1(N_1, N_2, N_3, V_{G3}, V_{G4}, V_{G6})$$
$$= E(N_1, N_2, N_3, V_{G3}, V_{G4}, V_{G6}) - E(N_1 - 1, N_2, N_3, V_{G3}, V_{G4}, V_{G6}),$$
$$\mu_2(N_1, N_2, N_3, V_{G3}, V_{G4}, V_{G6})$$
$$= E(N_1, N_2, N_3, V_{G3}, V_{G4}, V_{G6}) - E(N_1, N_2 - 1, N_3, V_{G3}, V_{G4}, V_{G6}),$$
$$\mu_3(N_1, N_2, N_3, V_{G3}, V_{G4}, V_{G6})$$
$$= E(N_1, N_2, N_3, V_{G3}, V_{G4}, V_{G6}) - E(N_1, N_2, N_3 - 1, V_{G3}, V_{G4}, V_{G6}).$$

Generally, the capacitively coupling between QD $i$ and gate $j$ allows the electron number $N_i$ changed by tuning the gate voltage $V_{Gj}$. The change in the gate voltage $\Delta V_{Gj-QDi}$ is found by solving the equation,

$$\mu_i(N_i, V_{Gj}) = \mu_i(N_i + 1, V_{Gj} + \Delta V_{Gj-QDi}).$$

We can find $\Delta V_{Gj-QDi}$ in terms of capacitances as

$$\Delta V_{G3-QD1} = \frac{e}{C_{G3-QD1} + \frac{C_3 C_{M12} + C_{M13} C_{M23}}{C_2 C_3 - C_{M23}^2} C_{G3-QD2}},$$

$$\Delta V_{G3-QD2} = \frac{e}{\frac{C_3 C_{M12} + C_{M13} C_{M23}}{C_1 C_3 - C_{M13}^2} C_{G3-QD1} + C_{G3-QD2}},$$

$$\Delta V_{G6-QD3} = \frac{e}{\frac{C_1 C_{M23} + C_{M12} C_{M13}}{C_1 C_2 - C_{M12}^2} C_{G6-QD2} + C_{G6-QD3}}.$$

From

$$\mu_1(N_1, N_2, N_3 + 1; V_{G3} + \Delta V_{G3-QD3}, V_{G6}) = \mu_1(N_1, N_2, N_3; V_{G3}, V_{G6}),$$

$$\mu_3(N_1 + 1, N_2, N_3; V_{G3}, V_{G6} + \Delta V_{G6-QD1}) = \mu_3(N_1, N_2, N_3; V_{G3}, V_{G6}),$$

$$\mu_3(N_1, N_2 + 1, N_3; V_{G3}, V_{G6} + \Delta V_{G6-QD2}) = \mu_3(N_1, N_2, N_3; V_{G3}, V_{G6}),$$

We obtain

$$\Delta V_{G3-QD3} = \frac{e}{\frac{C_2 C_3 - C_{M23}^2}{C_2 C_{M13} + C_{M12} C_{M23}} C_{G3-QD1} + \frac{C_3 C_{M12} + C_{M13} C_{M23}}{C_2 C_{M13} + C_{M12} C_{M23}} C_{G3-QD2}},$$

$$\Delta V_{G6-QD1} = \frac{e}{\frac{C_1 C_{M23} + C_{M12} C_{M13}}{C_2 C_{M13} + C_{M12} C_{M23}} C_{G6-QD2} + \frac{C_1 C_2 - C_{M12}^2}{C_2 C_{M13} + C_{M12} C_{M23}} C_{G6-QD3}},$$

$$\Delta V_{G6-QD2} = \frac{e}{C_{G6-QD2} + \frac{C_1 C_2 - C_{M12}^2}{C_1 C_{M23} + C_{M12} C_{M13}} C_{G6-QD3}}.$$

In our device, we have the following approximate conditions, $C_{M12}, C_{M13}, C_{M23} \ll C_{1,2,3}$, and the above equations can be simplified as

$$\Delta V_{G3-QD1} \approx \frac{e}{C_{G3-QD1}}, \tag{2}$$

$$\Delta V_{G3-QD2} \approx \frac{e}{C_{G3-QD2}}, \quad (3)$$

$$\Delta V_{G6-QD3} \approx \frac{e}{C_{G6-QD3}}, \quad (4)$$

$$\Delta V_{G3-QD3} \approx \frac{e}{\frac{C_2 C_3 - C_{M23}^2}{C_2 C_{M13} + C_{M12} C_{M23}} C_{G3-QD1}}, \quad (5)$$

$$\Delta V_{G6-QD1} \approx \frac{e}{\frac{C_1 C_2 - C_{M12}^2}{C_2 C_{M13} + C_{M12} C_{M23}} C_{G6-QD3}}, \quad (6)$$

$$\Delta V_{G6-QD2} \approx \frac{e}{\frac{C_1 C_2 - C_{M12}^2}{C_1 C_{M23} + C_{M12} C_{M13}} C_{G6-QD3}}. \quad (7)$$

In the simulations for the charge stability diagram shown in Figures 3a and 3c of the main article, we take the capacitance value of $C_{G4-QD2} \approx 5.3$ aF as extracted from the measurements of the DQD defined by gates G1 to G5. Because of the weak cross talk between gate G6 and QD2, the estimated gate capacitance value of $C_{G6-QD2} \approx 0.1$ aF is used. Capacitance $C_D$ is extracted from the stability diagram of the single QD in Figure S2 and capacitance $C_S$ is assumed to be the same as $C_D$. Other capacitance values are extracted from the stability diagrams shown in Figures 3a and 3c of the main article based on equations (2)-(7). The capacitance parameters used to obtain the stability diagrams shown in Figures 3b and 3d of the main article are given in Table S3.

**Table S3.** Capacitances used to simulate the charge stability diagrams of the TQD. The results of the simulations are shown in Figures 3b and 3d of the main article.

|  | $C_S$ (aF) | $C_D$ (aF) | $C_{G3-QD1}$ (aF) | $C_{G3-QD2}$ (aF) | $C_{G4-QD2}$ (aF) | $C_{G6-QD2}$ (aF) | $C_{G6-QD3}$ (aF) | $C_{M12}$ (aF) | $C_{M23}$ (aF) | $C_{M13}$ (aF) |
|---|---|---|---|---|---|---|---|---|---|---|
| **TQD in Figure 3b** | 6.2 | 6.2 | 2.5 | 2.6 | 5.3 | 0.1 | 5.5 | 0.8 | 2 | 0.7 |
| **TQD in Figure 3d** | 6.2 | 6.2 | 2.5 | 2.6 | 5.3 | 0.1 | 5.5 | 0.8 | 1.5 | 0.4 |